\documentclass[%
 reprint,
 amsmath,amssymb,
 aps,
floatfix
]{revtex4-2}

\usepackage{graphicx}
\usepackage{dcolumn}
\usepackage{bm}
\usepackage{braket}
\usepackage{comment}
\usepackage{hyperref}
\usepackage[nameinlink,poorman]{cleveref}

\crefname{equation}{Eq.}{Eqs.}
\crefname{figure}{Fig.}{Figs.}
\crefname{table}{Table}{Tables} 
\crefname{section}{Section}{Sections}
\crefname{chapter}{Chapter}{Chapters}
\crefname{appendix}{Appendix}{Appendices}
\crefname{algorithm}{Algorithm}{Algorithms}
\crefname{theorem}{Theorem}{Theorems}
\crefname{defn}{Definition}{Definitions}
\crefname{definition}{Definition}{Definitions}
\crefname{azm}{Assumption}{Assumptions}
\crefname{corollary}{Corollary}{Corollaries}
\crefname{lemma}{Lemma}{Lemmas}
\crefname{thmprop}{property}{properties}
\crefname{proposition}{Proposition}{Propositions}
\crefname{remark}{Remark}{Remarks}

\usepackage{tikz}
\usepackage{verbatim}

\begin{document}
\preprint{APS/123-QED}

\title{Exploiting biased noise in variational quantum models}





\author{Connor van Rossum}
 \email{c.vanrossum@uq.edu.au}

\author{Sally Shrapnel}%
\author{Riddhi Gupta}%
 \email{riddhi.gupta@uq.edu.au}
\affiliation{%
 School of Maths and Physics,
 University of Queensland,
 Brisbane, 4072 Australia
}%

\date{\today}

\begin{abstract}
Variational quantum algorithms (VQAs) have dominated literature as tools for demonstrating quantum utility on near-term quantum hardware, with applications in optimisation, quantum simulation, and machine learning. While researchers have studied how easy VQAs are to train, the effect of quantum noise on the classical optimisation process is still not well understood. Contrary to expectations, we find that twirling, which is commonly used in standard error-mitigation strategies to symmetrise noise, actually degrades performance in the variational setting, whereas preserving biased or non-unital noise can help classical optimisers find better solutions. Analytically, we study a universal quantum regression model and demonstrate that relatively uniform Pauli channels suppress gradient magnitudes and reduce expressivity, making optimisation more difficult. Conversely, asymmetric noise such as amplitude damping or biased Pauli channels introduces directional bias that can be exploited during optimisation. Numerical experiments on a variational eigensolver for the transverse-field Ising model confirm that non-unital noise yields lower-energy states compared to twirled noise. Finally, we show that coherent errors are fully mitigated by re-parameterisation. These findings challenge conventional noise-mitigation strategies and suggest that preserving noise biases may enhance VQA performance.
\end{abstract}

\maketitle


\section{Introduction}

Variational quantum algorithms (VQAs) feature predominantly in quantum applications development for noisy intermediate-scale quantum (NISQ) devices \cite{preskillQuantumComputingNISQ2018,cerezoVariationalQuantumAlgorithms2021,tillyVariationalQuantumEigensolver2022a}. Recent demonstrations for practical and near-term quantum utility on hardware have relied on VQAs in diverse fields such as combinatorial optimisation \cite{harriganQuantumApproximateOptimization2021, amaroCaseStudyVariational2022, duQuantumCircuitArchitecture2022b}, quantum simulations including chemistry \cite{kandalaHardwareefficientVariationalQuantum2017, kokailSelfVerifyingVariationalQuantum2019, aruteHartreeFockSuperconductingQubit2020, huangVariationalQuantumComputation2022, zhaoOrbitaloptimizedPaircorrelatedElectron2023, belalouiGroundStateEnergy2024, hanMultiLevelVariationalSpectroscopy2024}, and machine learning \cite{havlicekSupervisedLearningQuantum2019}. These hybrid quantum-classical algorithms employ parameterised quantum circuits whose parameters are trained to minimise a loss function, typically expressed as the expectation value of an observable. A persistent challenge for VQAs is ensuring that this loss-minimisation problem remains tractable for classical methods, particularly when quantum operations are subject to noise \cite{fontanaEvaluatingNoiseResilience2021}.

The trainability and noise-resilience of VQAs remain topics of active debate \cite{singkanipaUnitalNoiseVariational2025}. Large random VQAs are known to exhibit flat cost landscapes that hinder efficient training~\cite{mccleanBarrenPlateausQuantum2018,ragoneLieAlgebraicTheory2024}, while shallow random circuits may suffer from local minima or limited expressivity~\cite{anschuetzQuantumVariationalAlgorithms2022a,anschuetzUnifiedTheoryQuantum2025}. However, some of these trainability issues can be alleviated through innate problem structure \cite{hendersonQuantumAdvantageExponential2025}, \textit{a priori} information or non-random initialisation of variational parameters~\cite{parkHardwareefficientAnsatzBarren2024}. In such favourable regimes, noise-resilience of VQAs has relied on combating error via noise-aware compilation and control techniques~\cite{cincioMachineLearningNoiseresilient2021,duQuantumCircuitArchitecture2022} that optimise circuit architectures for specific hardware, as well as device-agnostic error mitigation strategies where dominant noise sources are associated with Clifford or measurement operations ~\cite{temmeErrorMitigationShortdepth2017a,bergProbabilisticErrorCancellation2023,guptaProbabilisticErrorCancellation2023}.  Typically, the first step in combating noise in quantum circuits is to symmetrise it via twirling, which reduces complex noise to more analytically convenient channels \cite{hashimRandomizedCompilingScalable2021,laydenTheoryQuantumError2025a}. Twirling over the Pauli group reduces noise to stochastic Pauli errors~\cite{durStandardFormsNoisy2005}, while twirling over the Clifford group further simplifies noise to a depolarising channel~\cite{emersonScalableNoiseEstimation2005,dankertExactApproximateUnitary2009}. However, despite its central role in state-of-the-art error mitigation protocols, it remains unclear how noise symmetrisation interacts with classical optimisation.

In this work, we show that noise symmetrisation can adversely affect VQA performance. Our results show a counterintuitive realisation: that noise symmetrisation and error mitigation for non-variational circuits cannot be naively applied to quantum circuits embedded in classical optimisation protocols. First, we analyse a family of VQAs where variational circuits are \textit{a priori} guaranteed to achieve zero loss under ideal conditions. Using asymptotically universal quantum models for classification and regression via data re-uploading circuits~\cite{schuldEffectDataEncoding2021a}, we introduce a channel-based framework to analyse the impact of noise on metrics for expressivity and trainability. Here, expressivity is quantified by the expected range of model outputs, while trainability is examined via the distribution of magnitudes of the loss gradients. Remarkably, we find that biased noise, such as amplitude damping, enables optimisers to steer models into regions of parameter space less affected by noise. This self-correcting behaviour is less accessible under more uniform Pauli noise profiles. We theoretically show that coherent unitary errors can be absorbed into the trainable parameters of the circuit, effectively reparameterising the model without degrading expressivity, consistent with prior observations that coherent errors are less detrimental than incoherent noise~\cite{wangNoiseinducedBarrenPlateaus2021a}. 

Our results extend from ideal zero-loss learning problems to general loss minimisation in variational quantum eigensolvers (VQEs). Using the transverse-field Ising model as a well-studied benchmark, we find circuits affected by amplitude damping perform better than those exposed to equivalent Pauli- or Clifford-twirled amplitude damping channels. Collectively, our results show that, contrary to its intended purpose in error mitigation, twirling can degrade variational performance by removing exploitable noise biases, hindering optimisation instead of improving it.

The structure of this document is as follows. In \cref{sec:theory_bg}, we introduce the theoretical setup and define the variational learning model that serves as our testbed. \Cref{sec:channel_representation} develops a channel-based representation of our variational model, enabling a systematic analysis of the effect of noise within the model. \Cref{sec:results} presents our main analytical and numerical results, including comparisons between biased, non-unital, and twirled noise channels. In \Cref{sec:discussion}, we discuss the broader implications of these findings for VQA design and noise-mitigation strategies. Finally, \Cref{sec:conclusion} concludes with a summary of our contributions and directions for future work.

\section{Theoretical Setup}
\label{sec:theory_bg}

To systematically characterise how biased or non-unital noise affects variational learning, we first establish a variational learning problem whose noiseless behaviour is analytically exact. While such analytical guarantees are rare for general VQAs, recent results in quantum machine learning provide them for a class of universal quantum models used in classification and regression tasks. These so‑called data re‑uploading circuits can learn any square‑integrable function~\cite{schuldEffectDataEncoding2021a}, offering a setting where the ideal, zero‑error solution is known. This makes them an ideal testbed for isolating the influence of biased and non‑unital noise from the inherent difficulty of the optimisation process. We will later relax these conditions and numerically characterise general VQA problems for which no analytical guarantees are available.

\begin{figure*}
\includegraphics[scale=0.9]{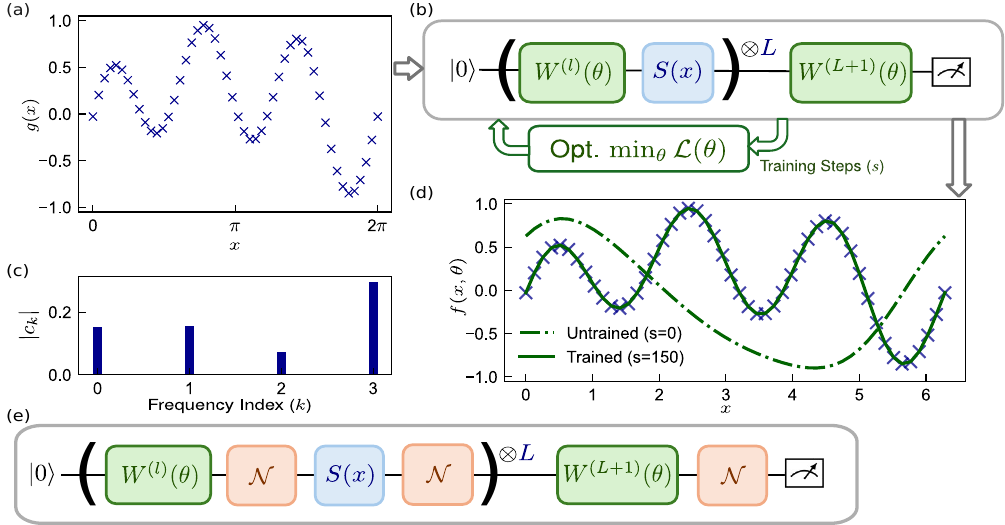}
\caption{\label{fig:circ_outs} 
(a) Sampled data points from a degree-three truncated Fourier series, used as the target function for supervised learning. 
(b) Ideal (noiseless) single-qubit data re-uploading circuit, where $S$ are data-encoding blocks and $W$ are trainable unitaries with parameters $\theta = \{\theta_1, \theta_2, \ldots\}$. 
(c) Fourier spectrum of the target function, showing the frequency components the model must learn. 
(d) Output of the quantum model, $f(x, \boldsymbol{\theta})$. With sufficient training, the model closely reproduces the target function. 
(e) Schematic of the noise model: quantum channels $\mathcal{N}$ are inserted after each $S$ and $W$ block to study the impact of different noise types on learning performance.}
\end{figure*}

A summary of the physical setup of data re-uploading circuits is shown in \cref{fig:circ_outs}. A key analytical convenience of these models is that their outputs admit a truncated Fourier series representation of the input data ~\cite{schuldEffectDataEncoding2021a,nemkovFourierExpansionVariational2023a}. The accessible frequency spectrum is determined solely by the eigenvalues of the data-encoding Hamiltonians, while the trainable unitaries and the measurement observable control the Fourier coefficients. This decomposition provides a natural lens for analysing expressivity and trainability. In particular, when the target function is expressible as a Fourier series, the ansatz circuit can be designed to ensure that a zero-loss solution exists in the noiseless regime. Consequently, any non-zero loss can be attributed to a combination of quantum noise and optimisation procedures. We therefore use these models as reliable testbeds to characterise the impact of biased and non-unital noise.

Restricting to the toy model of learning truncated Fourier series (\cref{fig:circ_outs}(a)), the structure of the noiseless circuit is illustrated in \cref{fig:circ_outs}(b). Each layer consists of a data-encoding gate $S(x)$ and a trainable unitary block $W(\boldsymbol{\theta})$, where,
\begin{equation}
    S(x) = e^{-ixH},
\end{equation}
where $x \in \mathbb{R}$ is classical input data and $H$ is a Hermitian operator. The trainable blocks $W(\boldsymbol{\theta})$ are general parameterised unitaries, often composed of single-qubit rotations and entangling gates. A full circuit with $L$ encoding–unitary layers is given by
\begin{equation}
    U(x, \boldsymbol{\theta}) = W^{(L+1)}(\boldsymbol{\theta}) S(x) W^{(L)}(\boldsymbol{\theta}) \cdots S(x) W^{(1)}(\boldsymbol{\theta}).
\end{equation}

It is well known that setting the degree of the Fourier series equal to the number of layers $L$ ensures the resulting quantum model can achieve perfect learning. This is depicted in \cref{fig:circ_outs}(d), where the trained model output $f(x, \boldsymbol{\theta})$ matches the target data (crosses) by minimising the loss $\mathcal{L}(\boldsymbol{\theta})$ with respect to tunable parameters $\boldsymbol{\theta}$. While the loss function can take various forms, we adopt the mean square loss,
\begin{align}
\mathcal{L}(\boldsymbol{\theta}) := \sum_x \frac{1}{2}\left(f(x,\boldsymbol{\theta})-g(x)\right)^2, \label{eqn:loss}
\end{align}
where $g(x)$ is the ground truth for data samples $x$. The quantum model output is the expectation value,
\begin{equation}
    f(x, \boldsymbol{\theta}) = \bra{0} U^\dag(x, \boldsymbol{\theta}) M U(x, \boldsymbol{\theta}) \ket{0},
\end{equation}
for initial state $\ket{0}$ and observable $M$. This admits a truncated Fourier series,
\begin{equation}
\label{eq:original_fourier_form}
    f(x, \boldsymbol{\theta}) = \sum_{\omega \in \Omega} c_\omega(\boldsymbol{\theta}) e^{i\omega x},
\end{equation}
where the frequency spectrum $\Omega$ is determined by the eigenvalues of $H$, and the coefficients $c_\omega$ depend on the trainable parameters $\boldsymbol{\theta}$ and the observable $M$.

The layered architecture of data re-uploading circuits also lends itself naturally to channel-based analysis. By inserting noise channels $\mathcal{N}$ after each $S$ and $W$ block, we model the noisy circuit as a composition of quantum channels, as illustrated in \cref{fig:circ_outs}(e). This structure enables a clean mapping to the Pauli transfer matrix (PTM) formalism, which we develop in \cref{sec:channel_representation} to analyse how different noise types contract the frequency spectrum and reshape the gradient landscape.

We employ two key properties to assess the practical utility of a variational quantum machine learning model: \textit{expressivity} and \textit{trainability}. Expressivity refers to the model's capacity to represent a broad class of functions or quantum states, often characterised via the Fourier spectrum of the output~\cite{schuldEffectDataEncoding2021a,nemkovFourierExpansionVariational2023a}. In the noiseless case, expressivity is commonly associated with the degree to which a quantum model uniformly explores the unitary group~\cite{holmesConnectingAnsatzExpressibility2022}. For noisy quantum computation, represented by the space of channels, we instead quantify expressivity by evaluating whether the quantum model can generate a sufficiently broad range of predictions to inform the loss calculation in \cref{eqn:loss}. Formally, we compute the expected maximum range of the model output,
\begin{align}
    \operatorname{Range}(\langle M \rangle) 
    &:= \mathbb{E}\Big[\max_{x} f(x,\boldsymbol{\theta}) - \min_{x} f(x,\boldsymbol{\theta})\Big], \label{eqn:expressivitymetric}
\end{align}
over an ensemble of randomly generated target functions $g'$ of fixed degree, scaled as,
\begin{equation}
    g(x) = a_{\max} g'(x) + b \quad \text{such that} \quad \exists \boldsymbol{\theta} : \mathcal{L}(\boldsymbol{\theta}) < \epsilon
\end{equation}
for some threshold $\epsilon\in \mathbb{R}$, and $a,b \in \mathbb{R}$. Here, the randomly generated target functions $g'$ are scaled by a factor $a$ to maximise their output range, such that there exists a parameter $\boldsymbol{\theta} \in \Theta$ for which the loss $\mathcal{L}(\boldsymbol{\theta})$ falls below the threshold $\epsilon$. Under ideal conditions, this range equals \(2\), reflecting the bounded spectrum of quantum observables. Since $f$ serves as the model prediction in \cref{eqn:loss}, its ability to approximate target functions is directly tied to the circuit's expressivity in both noise-free and noisy environments.

Meanwhile, trainability concerns the ability of classical optimisers to efficiently locate good parameters. It is governed by the structure of the loss landscape, particularly the magnitude and distribution of gradients \cite{wangNoiseinducedBarrenPlateaus2021a, holmesConnectingAnsatzExpressibility2022, wangCanErrorMitigation2024}. To this end, we characterise the distribution of gradient magnitudes of the model output,
\begin{align}
    \big|\partial f / \partial \theta_i \big|, \label{eqn:trainabilitymetric}
\end{align}
with respect to each trainable parameter \(\theta_i\). These gradients shape the \textit{loss landscape}, the high-dimensional surface defined by the loss function over parameter space, and govern the efficiency of gradient-based optimisation. The gradient landscape reveals key features such as barren plateaus~\cite{mccleanBarrenPlateausQuantum2018} and local minima~\cite{anschuetzQuantumVariationalAlgorithms2022a}. As shown in Ref.~\cite{okumuraFourierCoefficientParameterized2023b}, the sum of squared Fourier coefficients is exponentially suppressed under barren plateau conditions, linking spectral decay directly to vanishing gradients.

To compute these metrics analytically for data re-uploading circuits, we introduce a superoperator-based framework in \cref{sec:channel_representation} capturing the effect of representative noise models. Using this formalism, we analytically derive how different noise types alter the model’s output range and gradient landscape, providing insight into their impact on expressivity and trainability. We then validate these predictions through numerical simulations, revealing that noise with strong directional bias can sometimes be leveraged to maintain performance, whereas uniform noise consistently suppresses gradients and output range.

To demonstrate generality beyond quantum machine learning models, we additionally extend our numerical analysis to variational eigensolving problems. VQEs are hybrid quantum–classical algorithms designed to approximate the ground-state energy of a Hamiltonian $H$~\cite{peruzzoVariationalEigenvalueSolver2014}. In this setting, the circuit output corresponds to the expectation value of the Hamiltonian,
\begin{equation}
    f(\boldsymbol{\theta}) := \mathrm{Tr}\big(H \rho(\boldsymbol{\theta})\big),
\end{equation}
where $\boldsymbol{\theta}$ are trainable parameters and the observable is set to the Hamiltonian, $M \equiv H$. The variational principle guarantees that $f$ approaches the true ground-state energy $E_0$ from above, $f(\cdot) \geq E_0$. The optimisation task is therefore to minimise $f(\boldsymbol{\theta})$ over the parameter space, yielding an approximation to $E_0$. The loss function for VQE is simply,
\begin{equation}
    \mathcal{L}(\boldsymbol{\theta}) := f(\boldsymbol{\theta}),
\end{equation}
in contrast to the mean-squared loss considered earlier. Obtaining closed-form theoretical expressions to evaluate trainability metrics for VQEs using channel formalism appears intractable in general, other than limiting cases relying on Haar or Wishart-distributed random circuits~\cite{anschuetzUnifiedTheoryQuantum2025}. Meanwhile, extending the definition of expressibility from unitary VQAs to the space of quantum channels remains largely unexplored. The expressibility metric introduced in \cref{eqn:expressivitymetric} characterises the output range of $f$ ($f_{\max} - f_{\min}$), whereas VQE focuses on a single quantity, $f_{\min}$, and its deviation from $E_0$. Accordingly, we adopt the standard metric used in VQE literature and report the relative error between the obtained expectation value and $E_0$,
\begin{align}
    \text{Percentage Error} := \frac{f(\boldsymbol{\theta}) - E_0}{E_0}, \label{eq:percentageerror}
\end{align}
where the numerator is always positive since $f(\boldsymbol{\theta})$ approaches $E_0$ from above.

\section{Channel Representation of Noisy Learning}
\label{sec:channel_representation}

In this section, we introduce our analytical approach to isolating the impact of noise on classical loss minimisation. Using the PTM formalism and the analytical form of data-reuploading circuits, we isolate how each noise type attenuates or distorts the Fourier coefficients, thereby modifying the model’s output and gradient structure. We will find crucially that the frequency spectrum, determined by the eigenvalues of the encoding Hamiltonians, remains unchanged under noise. Only the amplitude and distribution of the coefficients vary, which directly affect the model’s ability to approximate target functions and to be trained efficiently.  Beyond data-reuploading circuits where analytical forms of quantum circuits are not directly useful to analysing the impact of noise models on loss minimisation, we instead provide numerical results for general variational loss minimisation problems in \cref{sec:results}.

We now formalise the PTM-based framework underpinning the analytical results in \cref{sec:results}. In this framework, quantum states are vectorised and their dynamics are described by linear maps acting on the operator space. Specifically, a density matrix \( \rho \) is vectorised as \( \ket{\rho} \rangle \), and its evolution under a noisy channel is described by,
\[
\ket{\rho} \rangle \longrightarrow \hat{\Lambda} \ket{\rho} \rangle,
\]
where \( \hat{\Lambda} \) is the superoperator corresponding to the full noisy circuit. The expectation value of an observable \( O \) is,
\begin{equation}
    \braket{O} = \operatorname{tr}(O\rho) = d \braket{\braket{O | \rho}}, 
\end{equation}
where \( \braket{\braket{A | B}} = \operatorname{tr}(A^\dag B)/d \) denotes the Hilbert–Schmidt inner product, and \( d \) is the Hilbert space dimension. In the PTM basis, both \( O \) and \( \rho \) are represented as vectors in Hilbert–Schmidt space. The dual vector \( \langle\bra{O} \) plays a role analogous to the Hermitian conjugate \( \bra{v} \) in standard Hilbert space. A detailed discussion of this dual interpretation is provided in \cref{appendix:superoperator_dual}.

We first present the ideal (noiseless) data re-uploading circuit, illustrated in \cref{fig:circ_outs}(b), within the PTM formalism. We define the model function \( f(x, \boldsymbol{\theta}) \) as the expectation value of an observable measured on a quantum state prepared by a parameterised circuit. In PTM notation, this is expressed as,
\begin{equation}
    f(x, \boldsymbol{\theta}) = d \braket{\braket{M|\hat{U}(x, \boldsymbol{\theta})|\rho_0}}, \label{eq:mexp}
\end{equation}
where \( \rho_0 \) is the initial state, \( M \) is the observable, and \( \hat{U}(x, \boldsymbol{\theta}) \) is the PTM representation of the full circuit. This expression follows from the superoperator dual formalism described in \cref{appendix:superoperator_dual}, where observables and states are treated as vectors in Hilbert–Schmidt space.

The circuit is assumed to have a layered structure, alternating between data-encoding and trainable unitary blocks,
\begin{equation}
    \hat{U}(x, \boldsymbol{\theta}) = \hat{W}^{(L+1)}(\boldsymbol{\theta}) \hat{S}(x) \cdots \hat{W}^{(2)}(\boldsymbol{\theta}) \hat{S}(x) \hat{W}^{(1)}(\boldsymbol{\theta}),
\end{equation}
where each \( \hat{W}^{(l)}(\boldsymbol{\theta}) \) is a trainable unitary block and \( \hat{S}(x) \) is the superoperator representation of a data-encoding gate of the form \( S(x) = e^{-ixH} \), with \( H \) a Hermitian generator.

In \cref{appendix:channel_derivations}, we show that for \( S(x) = e^{-ixH} \), the corresponding superoperator \( \hat{S}(x) \) takes the form \( V e^{ix\hat{\Sigma}} V^\dag \), where \( \hat{\Sigma} \) is a diagonal matrix whose entries correspond to the pairwise differences of the eigenvalues of \( H \), \( \hat{\lambda}_{j'j} = \lambda_{j'} - \lambda_j \). This decomposition isolates the data-dependent phase contributions and allows us to express the model output as a Fourier-like expansion,
\begin{equation} \label{eq:idealaform}
    f(x, \boldsymbol{\theta}) = d \sum_{\mathbf{j}} e^{ix\hat{\Lambda}_{\mathbf{j}}} \sum_{\mathbf{u}, \mathbf{v}}a_{\mathbf{j}, \mathbf{u}, \mathbf{v}}(\boldsymbol{\theta}),
\end{equation}
where $\mathbf{j}=j_1\cdots j_L$ indexes eigenvalues of the superoperator data encoding matrices $\hat{S}(x)$, and \( \hat{\Lambda}_{\mathbf{j}} = \hat{\lambda}_{j_1} + \cdots + \hat{\lambda}_{j_L} \). For full expressions of $a_{\mathbf{j}, \mathbf{u}, \mathbf{v}}$ and the gradient of $f(x,\boldsymbol{\theta})$, see \cref{appendix:ideal_model_derivations}. Grouping terms with equal $\hat{\Lambda}_{\mathbf{j}}=\omega$ recovers the Fourier form in \cref{eq:original_fourier_form},
\begin{equation}
    c_{\omega}(\boldsymbol{\theta}) = d \sum_{\substack{\mathbf{j} \\ \hat{\Lambda}_{\mathbf{j}} = \omega}} \sum_{\mathbf{u}, \mathbf{v}}a_{\mathbf{j}, \mathbf{u}, \mathbf{v}}(\boldsymbol{\theta}).
\end{equation}

Since the output \( f(x, \boldsymbol{\theta}) \) must be real-valued, the resulting function is necessarily a truncated Fourier series with symmetric frequency components. Additionally, there is a cumulative frequency \( \omega = 0 \in \Omega \), ensuring the presence of a constant term. 

We now extend the previous formalism to incorporate noise, enabling a detailed analysis of how different noise types affect the expressivity and trainability of data re-uploading quantum circuits. A noisy quantum circuit can be represented as a composition of alternating unitary and noise operations,
\begin{eqnarray}
    \hat{\Phi}(x, \boldsymbol{\theta}) &=& \hat{\mathcal{N}} \hat{W}^{(L+1)}(\boldsymbol{\theta}) \hat{\mathcal{N}} \hat{S}(x) \hat{\mathcal{N}} \hat{W}^{(L)}(\boldsymbol{\theta}) \cdots \nonumber \\
    && \hat{\mathcal{N}} \hat{W}^{(2)}(\boldsymbol{\theta}) \hat{\mathcal{N}} \hat{S}(x) \hat{\mathcal{N}} \hat{W}^{(1)}(\boldsymbol{\theta}), \label{eq:noisechan}
\end{eqnarray}
where \( \hat{\mathcal{N}} \) is the PTM of noise channel \( \mathcal{N} \), as illustrated in \cref{fig:circ_outs}(e). The noisy output is,
\begin{equation}
    \tilde{f}(x, \boldsymbol{\theta}) = d \braket{\braket{M|\hat{\Phi}(x, \boldsymbol{\theta})|\rho_0}}. \label{eq:noisef}
\end{equation}

\paragraph*{Pauli noise modified form.} The PTM representation \( \hat{\mathcal{N}} \) is diagonal with attenuation factors \( n_i \in [0,1] \), where $n_i$ are the Pauli eigenvalues. Through the same process as the ideal case, we then obtain,
\begin{equation}\label{eq:noisyfourier}
    \tilde{f}(x, \boldsymbol{\theta}) = d \sum_{\mathbf{j}} e^{ix\hat{\Lambda}_{\mathbf{j}}} \sum_{\mathbf{u}, \mathbf{v}} \tilde{a}_{\mathbf{j}, \mathbf{u}, \mathbf{v}}(\boldsymbol{\theta}),
\end{equation} 
see \cref{appendix:noisy_model_derivations} for the full form of modified coefficients $\tilde{a}_{\mathbf{j}, \mathbf{u}, \mathbf{v}}$ and the gradient of $\tilde{f}(x,\boldsymbol{\theta})$. The modified coefficients may be written with respect to the noiseless coefficients,
\begin{equation}\label{eq:pauliaforctors}
    \tilde{a}_{\mathbf{j}, \mathbf{u}, \mathbf{v}}(\boldsymbol{\theta}) = n_{\mathbf{u}} n_{\mathbf{v}} a_{\mathbf{j}, \mathbf{u}, \mathbf{v}}(\boldsymbol{\theta}),
\end{equation}
with \( n_{\mathbf{u}} = \prod_{l=1}^{L+1} n_{u_l} \) and \( n_{\mathbf{v}} = \prod_{l=1}^{L} n_{v_l} \) representing the cumulative attenuation across the respective Pauli eigenvalues.

To directly connect this formulation to the Fourier coefficients, we express the noisy Fourier coefficients as:
\begin{equation}\label{eq:noisecoeffrelation}
    \tilde{c}_\omega(\boldsymbol{\theta}) = n_\omega (\boldsymbol{\theta})c_\omega(\boldsymbol{\theta}),
\end{equation}
where the effective attenuation factor \( n_\omega \) is given by:
\begin{equation}
    n_\omega(\boldsymbol{\theta}) = \frac{\sum_{\substack{\mathbf{j} \\ \Lambda_{\mathbf{j}} = \omega}} \sum_{\mathbf{u}, \mathbf{v}}n_{\mathbf{u}} n_{\mathbf{v}} a_{\mathbf{j}, \mathbf{u}, \mathbf{v}}(\boldsymbol{\theta})}{\sum_{\substack{\mathbf{j}^{'} \\ \Lambda_{\mathbf{j}^{'}} = \omega}} \sum_{\mathbf{u}^{'}, \mathbf{v}^{'}}a_{\mathbf{j}^{'}, \mathbf{u}^{'}, \mathbf{v}^{'}}(\boldsymbol{\theta})}.
\end{equation}
Since $n_{\mathbf{u}} n_{\mathbf{v}} \in [0,1]$, therefore $n_\omega(\boldsymbol{\theta}) \in [0,1]$.

\paragraph*{Amplitude damping modified form.} Amplitude damping introduces a structural change: the index set expands to include substrings, reflecting qubit resets. The output retains a Fourier-like form,
\begin{equation}
    \tilde{f}(x, \boldsymbol{\theta}) = d \sum_{\mathbf{j'}} e^{ix\hat{\Lambda}_{\mathbf{j'}}} \sum_{\mathbf{u}, \mathbf{v}} \tilde{a}_{\mathbf{j'}, \mathbf{u}, \mathbf{v}}(\boldsymbol{\theta}), \label{eq:adform}
\end{equation}
where $\mathbf{j'}=j_k j_{k+1} \cdots j_L$ for $1 \leq k \leq L$ represents partial sequences after damping events.

\medskip
\noindent This is, to our knowledge, the first formulation that explicitly connects the Fourier structure of data re-uploading circuits with the linear action of superoperators in the PTM basis. It provides a rigorous foundation for analysing how different noise types reshape Fourier coefficients and the geometry of the gradient landscape. 

For simulated noise models, amplitude damping noise represents a decay from the excited state $\ket{1}$ to the ground state $\ket{0}$ with probability $\gamma$, which we refer to as the noise strength in our figures. We apply Pauli and Clifford twirling directly to the amplitude damping channel, in lieu of approximately twirling noisy non-Clifford gates directly. For the Pauli-twirled amplitude damping channel, we construct an equivalent Pauli noise model by solving for Pauli error probabilities $p_X, p_Y, p_Z$ such that the diagonal elements of the PTM of this Pauli channel match those of the amplitude damping channel. This yields:
\[
p_X = p_Y = \frac{\gamma}{4}, \qquad
p_Z = \frac{2 - \gamma - 2\sqrt{1 - \gamma}}{4}.
\]
The resulting Pauli error distribution is:
\[
\{ ('I', 1 - 2p_X - p_Z),\; ('X', p_X),\; ('Y', p_Y),\; ('Z', p_Z) \}.
\]

For the Clifford-twirled case, noise is reshaped into a depolarising channel. The PTM of a depolarising channel is diagonal with elements $\{1, 1-p_{\text{depol}}, 1-p_{\text{depol}}, 1-p_{\text{depol}}\}$ in the one qubit case, where $p_{\text{depol}}$ is the depolarising rate. To match a Clifford-twirled amplitude damping channel, we set $1-p_{\text{depol}}$ equal to the average of the non-identity diagonal elements of the amplitude damping PTM. This gives:
\[
p_{\text{depol}} = \frac{\gamma + 2 - 2\sqrt{1 - \gamma}}{3}.
\]
Details on how twirling affects the PTM representations of single-qubit channels are provided in \cref{appendix:channeltwirling}.


\section{Results \label{sec:results}}
In this section, we investigate how different noise models influence the performance of VQAs. First, we theoretically analyse data re-uploading circuits using the formalism of quantum channels represented as PTMs. Here, we obtain expressions for expressivity and trainability of data re-uploading circuits when input data represents Fourier signals. We then extend numerical investigations to VQE, where we choose $H$ to represent the Hamiltonian of a transverse-field 2D Ising model with periodic boundary conditions, a common benchmark for VQE studies~\cite{cerezoVariationalQuantumAlgorithms2021}. In all cases, the noiseless regime is contrasted with four noise models: coherent errors, amplitude damping, and Pauli channels equivalent to Pauli- or Clifford-twirled amplitude damping. For the latter two noise models, we sample VQAs subject to Pauli channels that represent a Pauli- or Clifford- twirled amplitude damping channel, instead of approximately twirling noisy non-Clifford gates \cite{laydenTheoryQuantumError2025a}. Our goal is to compare noise models with increasing levels of symmetrisation, from Pauli-twirl to Clifford-twirl, applied to the same amplitude damping channel.

\subsection{VQAs for Machine Learning}

We first present a theory result that applies to asymptotically universal quantum models for classification and regression in quantum machine learning. While their Fourier structure in the noiseless regime is well known, we reformulate this description using the superoperator formalism, reproducing \cref{eq:original_fourier_form} via channel representations. As detailed in \cref{sec:channel_representation}, we extend this channel representation to then include noise that previously could not be analysed using unitary representations alone. A key insight emerges: noise alters the magnitude and distribution of Fourier coefficients but not the frequency spectrum, which remains determined by the data-encoding Hamiltonians. Leveraging this property, we systematically analyse how representative noise channels reshape the output range and gradient landscape, thereby impacting expressivity and trainability.

\paragraph*{Impact of noise on expressivity.} We analyse expressivity via \cref{eqn:expressivitymetric} by characterising the impact of different noise models on the output function. For Pauli and coherent noise, the impact on the output function manifests as altered Fourier coefficients in \cref{eq:original_fourier_form}, while amplitude damping introduces additive terms that preclude a simple closed-form expression. When noise channels are composed with the channel representation of the data re-uploading circuit, the resulting noisy output function can be expressed as
\begin{equation}
\tilde{f}(x, \boldsymbol{\theta}) = \sum_{\omega \in \Omega} \tilde{c}_{\omega}(\boldsymbol{\theta}) e^{i\omega x}, \label{eq:noisyfourierform}
\end{equation}
which retains the structure of \cref{eq:original_fourier_form}, with noise effects isolated in $\tilde{c}_{\omega}(\boldsymbol{\theta})$, and $\Omega$ is unchanged. For noise diagonal in the PTM representation, such as Pauli noise, the circuit output inherits multiplicative factors $\in [0,1]$ that scale the Fourier coefficients. These factors correspond to the diagonal elements of the PTM, which are the Pauli eigenvalues, and their effect is to contract the output function toward zero, thereby reducing expressivity. As per \cref{eq:noisecoeffrelation}, these coefficients can be recast in the simpler form,
\[
\tilde{c}_{\omega}(\boldsymbol{\theta}) = n_{\omega}(\boldsymbol{\theta}) c_{\omega}(\boldsymbol{\theta}),
\]
where \( 0 \leq n_{\omega}(\boldsymbol{\theta}) \leq 1 \) quantifies the attenuation. 

In contrast, amplitude damping introduces not only multiplicative attenuation but also a partial reduction in the effective circuit depth. Prior work has shown that this noise mechanism truncates circuit evolution, thereby limiting the achievable depth of quantum circuits~\cite{meleNoiseinducedShallowCircuits2024}. For qubits affected by damping, the channel effectively resets them to the ground state, erasing earlier operations and enabling subsequent layers to act on a simplified state. This depth reduction enables damped qubits to still contribute meaningfully to lower-frequency components in the Fourier spectrum, which helps preserve the model’s representational capacity. Consequently, although the multiplicative attenuation of Fourier coefficients under amplitude damping mirrors that observed with Pauli noise, there are accompanying terms which additionally contribute to the lower frequency Fourier coefficients, as shown by \cref{eq:adform}. Finally, we remark that coherent noise corresponds to unitary errors. Since the set of unitaries is closed under composition, coherent noise can be absorbed into the parameterisation of the model. For each \( \boldsymbol{\theta} \), there exists a reparameterised \( \boldsymbol{\theta}' \) such that $\tilde{c}_{\omega}(\boldsymbol{\theta}) \equiv c_{\omega}(\boldsymbol{\theta}')$. Our theoretical results imply that coherent noise preserves expressivity.

\paragraph*{Impact of noise on trainability.} To analyse trainability, we consider the gradient of the model output with respect to each parameter. Traditionally, these metrics have been used in literature to identify regimes where gradients vanish or concentrate, and we numerically examine distributions of gradient magnitudes in \cref{eqn:trainabilitymetric} under different noise models. For Pauli noise, loss gradients are scaled by the same attenuation factors that affect the model output, leading to flatter loss landscapes and diminished gradient magnitudes,
\begin{align*}
\frac{\partial \tilde{f}}{\partial \theta_i}(x, \boldsymbol{\theta}) = \sum_{\omega} \frac{\partial \tilde{c}_\omega}{\partial \theta_i}(\boldsymbol{\theta}) e^{i\omega x}, \quad \frac{\partial \tilde{c}_\omega}{\partial \theta_i}(\boldsymbol{\theta}) = n_{\omega}(\boldsymbol{\theta}) \frac{\partial c_\omega}{\partial \theta_i}(\boldsymbol{\theta}). 
\end{align*} 
For coherent noise, if \( h \) maps \( \boldsymbol{\theta} \mapsto \boldsymbol{\theta}' \), then,
\[
\frac{\partial \tilde{f}}{\partial \theta_i}(x, \boldsymbol{\theta}) = \frac{\partial f}{\partial \theta_i} \big( x, h(\boldsymbol{\theta}) \big) \frac{\partial h}{\partial \theta_i}(\boldsymbol{\theta}),
\]
suggesting that coherent noise may preserve gradient structure. As with expressivity metrics for amplitude damping, we find loss gradients appear to be suppressed with terms matching Pauli noise, however, again, there are additional terms corresponding to shorter depth circuits contributing to lower frequency coefficient gradients. Further details on gradient expressions for amplitude damping are provided in \cref{appendix:A}.

In summary, our extension of Fourier-based analysis of data re-uploading circuits to include realistic noise models reveals several insights. Noise processes that are diagonal in the PTM basis, such as Pauli or depolarising noise, attenuate Fourier coefficients, shrinking the output range and flattening the loss landscape. Amplitude damping introduces both multiplicative attenuation and additive bias. For this non-unital noise, multiplicative attenuation factors resemble those of diagonal (Pauli) noise channels, while additive bias contributes to the lower frequency Fourier coefficients. Meanwhile, we expect coherent noise to always be absorbed into reparameterised unitaries.

These theoretical insights provide a principled understanding of how different noise types influence the learnability of quantum models. In particular, they highlight the distinct ways in which noise can degrade expressivity by suppressing the magnitude of Fourier coefficients and affect trainability by flattening or distorting the loss landscape. These effects vary by noise type: Pauli noise causes uniform contraction, amplitude damping introduces biases, and coherent noise can be absorbed but may complicate optimisation.

\begin{figure}[]
\includegraphics[]{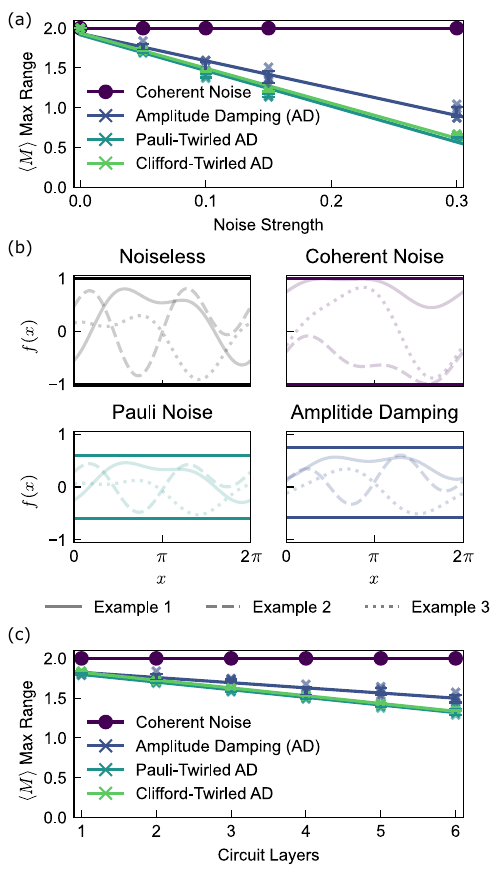}
\caption{\label{fig:exprange} 
Impact of noise on model expressivity. 
(a) Output range of a single-qubit data re-uploading circuit under different noise channels, determined by increasing the amplitude of target functions until the model fails to fit the data. Error bars represent the mean and standard deviation across multiple samples, and lines of best fit were computed using linear regression. The comparison highlights the difference between biased noise (amplitude damping) and more uniform noise (Pauli noise, here simulated as the equivalent of Pauli- and Clifford-twirled amplitude damping), with the former preserving a broader output range and enabling more expressive behaviour. Coherent unitary errors, by contrast, show no degradation in expressivity, as they can be absorbed into the trainable parameters.
(b) Upper and lower bounds of the model's output under different noise conditions. The same three example circuits are evaluated across multiple noise types to illustrate how each type deforms the achievable output range.
(c) Output range comparison across circuits of increasing depth, where the degree of the target function matches the number of layers. As in (a), error bars denote the mean and standard deviation, and lines of best fit were obtained via linear regression. Results show that deeper circuits amplify the impact of noise, with more uniform channels causing more severe degradation.}
\end{figure}

Having established analytically how different noise models influence expressivity and trainability, we now turn to numerical simulations to validate these predictions. Our objective is to quantify the practical impact of noise on data re-uploading circuits and compare these with our analytical expectations. We directly match the degree of target Fourier functions with the number of layers $L$ to ensure model error is zero \cite{schuldEffectDataEncoding2021a}. To reduce the risk of miscalibrated optimisation protocols, all hyperparameters were selected via Bayesian optimisation targeting convergence on representative target Fourier functions; see \cref{appendix:B}.

We numerically quantify expressivity under noise in \cref{fig:exprange}. Here, we measure the output range of the trained model under different noise conditions. For each seeded target function, we use binary search to find the largest output range for which the model converges as per \cref{eqn:expressivitymetric}. A trial succeeds if the final loss is below \(\epsilon = 10^{-5}\) within the training budget of 150 steps. \cref{fig:exprange}(a) shows the maximum range for which the model can successfully learn a target function under coherent noise, amplitude damping, and channels equivalent to Pauli-/Clifford-twirled amplitude damping, where we use the same set of seeded functions for each noise type. The twirled channels are simulated as equivalent noise channels rather than physically twirled circuits, due to the non-Clifford nature of the ansatz. 

Our results confirm our theoretical analysis, showing that coherent noise (purple) can be absorbed into the learned parameters and has no effect on the function range. In contrast, incoherent noise channels reduce the model's expressivity by contracting the output range. Our comparison of amplitude damping with its Pauli-twirled and Clifford-twirled counterparts reflects an increase in noise symmetrisation. These Pauli channels show a reduction in expressivity at a rate consistent with our theoretical analysis. Meanwhile, amplitude damping also reduces expressivity but preserves a broader range of model outputs. This preservation suggests that the additional terms in our theoretical analysis, linked to its directional bias, are beneficial to the model.

\begin{figure*}
\includegraphics{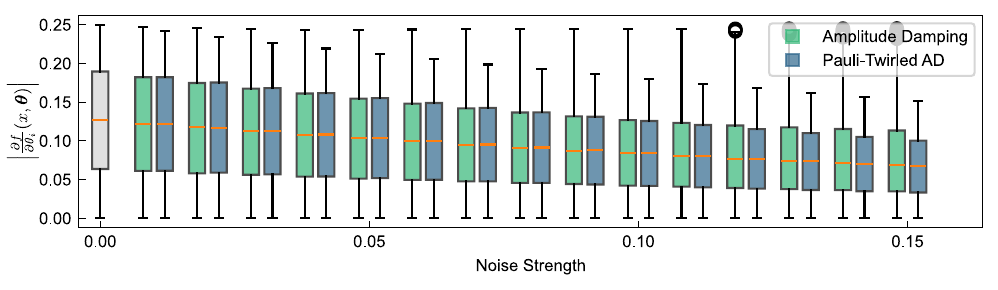}
\caption{\label{fig:grad_decay}
Effect of noise on gradient magnitudes during training. 
Box plots show distributions of absolute gradients \(\left| \frac{\partial \tilde{f}}{\partial \theta_i} \right|\) for a two-layer data re-uploading circuit under varying strengths of amplitude damping and Pauli-twirled amplitude damping. Each distribution uses 10{,}000 random parameter sets and inputs. Pauli-twirled noise causes a sharp decay in gradient magnitudes, indicating reduced trainability, while amplitude damping preserves a richer gradient profile, enabling more effective optimisation.}
\end{figure*}

To assess trainability, we characterise the distribution of gradient magnitudes under noise (\cref{fig:grad_decay}). Having seen that coherent noise can be addressed via reparameterisation, we focus only on amplitude damping noise and compare this to the equivalent Pauli channel obtained via Pauli twirling, with analogous results for Clifford twirling provided in \cref{appendix:pauli_loss}. For each of 10{,}000 randomly sampled parameter sets \(\boldsymbol{\theta} = \{\theta_1, \ldots, \theta_9\}\), we randomly select one parameter \(\theta_i\) and one input \( x \sim \mathcal{U}(0, 2\pi) \). Gradients \(\left| \frac{\partial \tilde{f}}{\partial \theta_i}(x, \boldsymbol{\theta}) \right|\) are computed analytically for each noise setting, and these gradient distributions are plotted as boxplots in \cref{fig:grad_decay}. We observe that Pauli-twirled noise consistently suppresses gradient magnitudes, leading to flatter loss landscapes and reduced sensitivity to parameter updates. In contrast, amplitude damping retains a richer gradient distribution. This suggests that amplitude damping preserves some trainability, even as it reduces average gradient values. Gradient distributions use the same ensemble of random target functions across all noise conditions to ensure consistency.

These numerical investigations suggest that incoherent noise degrades both expressivity and trainability, while coherent noise appears almost entirely addressable via reparameterisation. Crucially, we find that symmetrising noise through twirling can significantly worsen performance. 

\subsection{VQAs for Variational Eigensolving}
\begin{figure}[]
\includegraphics[]{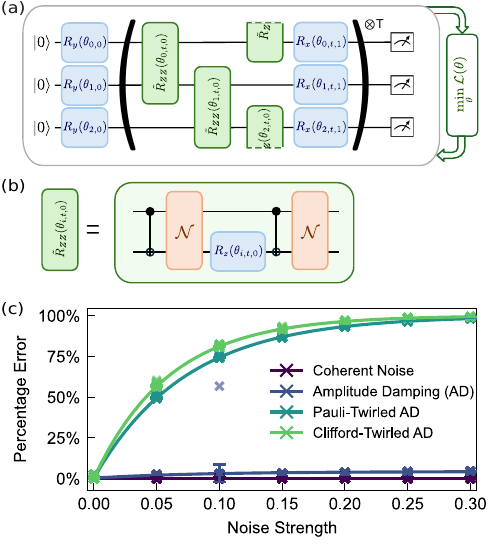}
\caption{\label{fig:VQE_error}
Overview of the VQE setup and the impact of noise on solution quality. 
(a) Parameterised VQE circuit with $T$ Trotter steps, where $\boldsymbol{\theta}$ denotes the set of trainable parameters and $t$ indexes the individual Trotter steps. The circuit is optimised to minimise the loss function $\mathcal{L}(\boldsymbol{\theta}) = \mathrm{Tr}(H \rho(\boldsymbol{\theta}))$, where $\rho(\boldsymbol{\theta})$ is the pre-measurement density matrix prepared by the circuit and $H$ is the problem Hamiltonian.
(b) Implementation of the $R_{ZZ}$ gate acting on qubits $i$ and $i+1$, along with the noise injection strategy, where noise is introduced following each two-qubit gate (CNOT).
(c) Effect of noise on VQE solutions for a transverse-field 2D Ising model with periodic boundary conditions. As noise strength increases, solutions deviate from the true ground-state energy ($E_0 = -3.232$) and approach zero. Error bars, shown for all data points, represent the mean and standard deviation of the relative error across 100 samples; most are too small to be visible at this scale. Coherent noise (modelled via a controlled-$Y$ rotation) is absorbed into the trainable parameters, showing no degradation. Amplitude damping approximates the true solution, with one outlier where the optimiser was trapped in a local minimum. Pauli- and Clifford-twirled amplitude damping exhibit exponential decay towards zero, indicating severe quality loss under twirling. Lines of best fit were obtained using either linear regression (for coherent noise) or non-linear curve fitting to an asymptotic decay model of the form $y = a e^{-b x} + c$ (for decoherence-based noise channels).}
\end{figure}

We now move beyond analytically tractable VQAs to VQAs that variationally prepare lowest-energy eigenstates and approximate the ground-state energy of a Hamiltonian $H$~\cite{peruzzoVariationalEigenvalueSolver2014}. To benchmark against well-studied eigensolving problems, we focus on the transverse-field Ising model for three qubits, using the Hamiltonian,
\begin{align}
H = -J\sum_i Z_i Z_{i+1} - h\sum_i X_i,
\end{align}
with $J = 1.0$ and $h = 0.5$. The overall setup is illustrated in \cref{fig:VQE_error}(a) as a parameterised circuit. Each qubit $i$ is initialised with a tunable $Y$-rotation $\theta_{i,0}$, followed by four Trotter step blocks. Each block $t$ applies $R_{ZZ}(\theta_{i,t,0})$ gates, under periodic boundary conditions, where nearest-neighbour qubits $i$ and $i+1$ are entangled. Additionally, each block, applies a parameterised $X$-rotation $\theta_{i,t,1}$ to each qubit. Noise is injected after each two-qubit gate, as illustrated in \cref{fig:VQE_error}(b), assuming single-qubit gates are noiseless, consistent with common error-mitigation assumptions~\cite{bergProbabilisticErrorCancellation2023}.

We report numerical results showing the performance impact on VQEs when two-qubit gate noise is symmetrised in \cref{fig:VQE_error}(c). Here, the true energy value $E_0 = -3.232$ is obtained by exact diagonalisation of $H$ for this toy problem. The $y$-axis reports percentage error (\cref{eq:percentageerror}) between estimated energy values and $E_0$ under varying noise strengths. As the noise strength increases, the quality of the VQE solutions deteriorates, with the estimated energies drifting away from the true ground-state value and approaching zero. This degradation is highly dependent on the noise structure. Consistent with our expressivity results, coherent noise shows no observable degradation, as its effects are absorbed into the trainable parameters. Amplitude damping generally maintains strong performance, where the lone outlier in \cref{fig:VQE_error}(c) (`x' marker) for noise-strength $0.1$ corresponds to a single optimiser failure out of 100 trials. In contrast, both Pauli-twirled and Clifford-twirled versions of amplitude damping exhibit exponential decay in solution quality. Again, these twirled channels are simulated as equivalent noise channels rather than physically twirled circuits, due to the non-Clifford nature of the ansatz. This behaviour mirrors the contraction of output range observed in \cref{fig:exprange} and reinforces the conclusion that symmetrisation can amplify the most disruptive components of noise for a given circuit structure. Full configuration details for the optimisation procedure are provided in \cref{appendix:sim_details}.

\section{Discussion \label{sec:discussion}}

\noindent Our simulations demonstrate that different noise types impact variational quantum circuits in fundamentally distinct ways. Coherent noise can be absorbed into the model’s parameterisation with minimal impact on the quantum model's expressivity or gradient magnitudes. Incoherent noise degrades both expressivity and trainability metrics. However, amplitude damping, while still harmful, better preserves expressivity and gradient magnitudes compared to an equivalent Pauli channel obtained from Pauli twirling. In particular, twirling coherent errors would similarly reduce them to Pauli noise, eliminating the possibility of reparameterisation to absorb the error.

Our analysis relies on several simplifying assumptions. First, we treat noise channels and circuit blocks as separable, modelling the circuit as alternating unitaries and noise superoperators. In practice, noise may act within blocks and interact non-trivially with parameterised gates, particularly within trainable blocks composed of multiple gates. Second, we assume each \( W^{(l)} \) is an arbitrary unitary. While this is a common theoretical simplification, real-world devices often restrict the parameter space to a subset of the full unitary group. In such cases, coherent errors may push the circuit outside the expressible subspace, potentially leading to more severe degradation than observed here. Third, we compare VQAs under non-unital noise to those under Pauli or depolarising channels that would arise if we could, in fact, perfectly twirl non-Clifford operations. While some non-Clifford operations can be decomposed into layers that permit twirling of dominant noise sources, known protocols for twirling noisy non-Clifford gates generally only perform averaging over commuting subsets of the twirling group \cite{laydenTheoryQuantumError2025a}, yielding block-diagonal PTMs. Even if dominant noise sources can be twirled, our idealised approach avoids the sampling overhead of repeated twirls required for average noise channels to be approximately diagonal. Relaxing this assumption to explicitly twirl noisy non-Clifford gates with almost-diagonal PTMs is the subject of future work.

Despite these limitations, we expect our findings to generalise to a broad class of variational quantum circuits. Although our simulations focused on single-qubit circuits, the underlying mathematical framework is fully general and applies to multi-qubit systems. Moreover, the observed trends in gradient suppression and expressivity loss are consistent with known effects of decoherence in quantum systems, suggesting that our conclusions extend beyond the specific setting studied here. We emphasise that our analysis does not address the generalisability of variational quantum models. The data re-uploading setting deliberately uses target functions that are perfectly representable by the ansatz in the absence of noise, allowing us to isolate the effects of noise on expressivity and trainability. As a consequence, our results speak to optimisation behaviour rather than a model’s ability to generalise to unseen data or perform out-of-distribution prediction.

Finally, we note that while exploiting biased noise is well-established in quantum error correction \cite{aliferisFaulttolerantQuantumComputation2008,tuckettFaulttolerantThresholdsSurface2020}, the implications for VQAs are less understood. Recent work has shown that non-unital noise can suppress entropy and prevent barren plateaus by effectively truncating deep circuits to logarithmic depth~\cite{meleNoiseinducedShallowCircuits2024}. This entropy suppression aligns with our observation that amplitude damping preserves expressivity and gradient magnitudes better than symmetrised Pauli noise. One implication of our results is that noise-tailoring techniques may inadvertently worsen the performance of parameterised quantum circuits. While we focused on twirled amplitude noise in the main text, \cref{appendix:asym_pauli} explores biased Pauli channels and how ansatz design can amplify or reduce the role of biased noise in improving overall performance.

To further probe the role of circuit design in exploiting noise structure, we consider a non-physical reversed amplitude damping channel in \cref{appendix:VQE_ad_reverse}, which excites $\ket{0}$ to $\ket{1}$. Although unphysical, this variant highlights the circuit-level dependence on noise directionality: reversed damping performs significantly worse than standard amplitude damping, yet still outperforms its Pauli- and Clifford-twirled counterparts. These results suggest that bias-preserving circuit design could enhance optimisation under non-unital noise. A detailed analysis of circuit architectures designed to exploit self-correcting behaviour under biased or non-unital noise is an exciting direction for future work.

\section{Conclusion \label{sec:conclusion}}

Our work counterintuitively indicates that bias-preserving noisy variational circuits may improve overall performance, whereas naively twirling noise may instead hinder classical optimisation of VQAs. We studied two families of VQAs with broad benchmarking appeal: data re-uploading circuits as universal models for classification and regression, and variational eigensolvers for a perturbed Ising model. For the former, we extended Fourier-based analysis to noisy settings using channel formalism, showing that incoherent noise suppresses Fourier coefficients and gradients without altering the frequency spectrum. These effects were confirmed through simulation, which showed that more uniform noise leads to more severe degradation in both output range and gradient distribution. In the latter case, our numerical investigations of VQE confirmed that biased noise profiles yield better variationally prepared ground states. This performance differential is attributed to the earlier predictions that biased noise profiles preserve a richer gradient profile and retain a greater level of expressivity. In all cases, coherent noise, by contrast, was largely compensatable through parameter reconfiguration.

Our findings suggest VQAs may benefit from retaining biases in intrinsic noise to preserve expressivity and trainability of quantum models. One implication appears to be that commonly used noise reshaping techniques, such as Pauli twirling, may inadvertently degrade the performance of variational algorithms. This perspective opens new avenues for designing noise-resilient quantum models that are better suited to the operational realities of NISQ hardware. Further research is also required in articulating expressivity and trainability metrics for VQAs in noisy settings. In particular, the development of new metrics could help delineate how much error mitigation is outsourced to a classical optimiser and what residual noise is in-scope for an error mitigation protocol to actually address. Finally, our work represents a general formalism that can be used to study the scalability of these effects with increased circuit depth and qubit counts.


\section{Acknowledgements}
C.v led the development of theoretical frameworks; performed all simulations, developed software tools, managed all data and analysis. Both S.S. and R.S.G co-supervised the project while R.S.G. set the original research direction. All authors contributed to technical analysis of results and co-wrote the manuscript.

C.v. is supported by the University of Queensland's Graduate School, the Queensland Government Department of Environment, Science and Innovation, and the ARC Centre of Excellence for Engineered Quantum Systems (CE17010000). S.S. is supported by the ARC Centre of Excellence for Engineered Quantum Systems (CE17010000). R.S.G. is supported by UQ’s Queensland Digital Health Center via funding from UQ’s Health Research Accelerator (HERA) initiative.

All authors declare no financial or non-financial competing interests.

All scripts and data are available upon reasonable request and will be made publicly available at time of publishing.

\bibliography{Paper1}

\appendix
\crefalias{section}{appendix} 
\crefalias{subsection}{appendix}
\crefalias{subsubsection}{appendix}
\section{Theoretical Background and Additional Proofs}
\label{appendix:A}
\subsection{Superoperator Dual and Expectation Values}
\label{appendix:superoperator_dual}

It is often useful to represent a quantum channel \( \Lambda \) as a superoperator \( \hat{\Lambda} \) acting on vectorised density matrices:
\begin{equation}
    \ket{\rho}\rangle \rightarrow \ket{\rho'}\rangle = \hat{\Lambda} \ket{\rho}\rangle,
\end{equation}
where \( \ket{\rho}\rangle \) is the vectorisation of the density matrix \( \rho \). A \( d \times d \) density operator in a \( d \)-dimensional Hilbert space becomes a \( d^2 \)-dimensional vector in Hilbert–Schmidt space, and \( \hat{\Lambda} \) is a \( d^2 \times d^2 \) matrix acting in that space.

The vectorisation of an operator \( O \) depends on the chosen basis \( \{ \ket{k} \rangle \} \) in Hilbert–Schmidt space:
\begin{equation}
    \ket{O}\rangle = \sum_k \ket{k}\rangle \langle\braket{k|O}\rangle, \label{eq:innerpvec_appendix}
\end{equation}
where \( \langle\braket{A|B}\rangle = \operatorname{tr}(A^\dag B)/d \) is the Hilbert–Schmidt inner product. The matrix elements of the superoperator are then given by:
\begin{equation}
    \hat{\Lambda}_{ij} = \langle\braket{k_i|\Lambda(k_j)}\rangle. 
\end{equation}

In the Pauli Transfer Matrix (PTM) representation, we choose the basis \( \{ \ket{k} \rangle \} = \{ P_k \} \), where \( P_k \) are Pauli operators. The PTM elements are:
\begin{equation}
    \hat{R}_{ij} = \frac{1}{d} \operatorname{tr}(P_i \Lambda(P_j)).
\end{equation}

Using the completeness of the Pauli basis, any operator can be expanded as:
\begin{equation}
    O = \sum_k \frac{1}{d} \operatorname{tr}(P_k O) P_k.
\end{equation}
Substituting this into the trace expression for expectation:
\begin{eqnarray}
    \braket{O} &=& \operatorname{tr}(O\rho) \nonumber \\
    &=& \frac{1}{d^2} \operatorname{tr} \left( \left( \sum_j P_j \operatorname{tr}(P_j O) \right) \left( \sum_k P_k \operatorname{tr}(P_k \rho) \right) \right) \nonumber \\
    &=& \frac{1}{d^2} \sum_{j,k} \operatorname{tr}(P_j O) \operatorname{tr}(P_k \rho) \operatorname{tr}(P_j P_k). \label{eq:trace_expansion}
\end{eqnarray}

We may expand higher dimensional Pauli operators into a Kronecker product of base Pauli's:
\begin{eqnarray}
    \operatorname{tr}(P_j P_k) &=& \operatorname{tr}\left((P_{j_1} \otimes \cdots \otimes P_{j_n})(P_{k_1} \otimes \cdots \otimes P_{k_n})\right) \nonumber \\
    &=& \prod_{l=1}^n \operatorname{tr}(P_{j_l} P_{k_l}) = d \, \delta_{j_1, k_1} \cdots \delta_{j_n, k_n}.
\end{eqnarray}

By Eq.~\eqref{eq:innerpvec_appendix}, \( \operatorname{tr}(P_k O) = d [\ket{O}\rangle]_k \), where \( [\ket{O}\rangle]_k \) denotes the \( k \)-th component of the vectorised operator. Substituting into Eq.~\eqref{eq:trace_expansion}, we obtain:
\begin{eqnarray}
    \braket{O} &=& \frac{1}{d} \sum_j \operatorname{tr}(P_j O) \operatorname{tr}(P_j \rho) \nonumber \\
    &=& d \sum_j [\ket{O}\rangle]_j [\ket{\rho}\rangle]_j. \label{eq:expelewise}
\end{eqnarray}

Combining \eqref{eq:expelewise} with the inner-product representation of an observable's expectation, we see that the dual vector \( \langle\bra{O} \) in Hilbert–Schmidt space plays a role analogous to the Hermitian conjugate \( \bra{v} \) in Hilbert space. Specifically, \( \langle\bra{O} \) is the row vector whose components are the complex conjugates of those in \( \ket{O}\rangle \), and the expectation value is given by their inner product:
\[
\braket{O} = d \langle\bra{O} \ket{\rho}\rangle.
\]

\subsection{Derivations of Channel Representations}
\label{appendix:channel_derivations}

We provide here the full derivation of the channel representations used in \cref{sec:channel_representation}, including the Liouville superoperator form and the Pauli Transfer Matrix (PTM) transformation.

\subsubsection{Liouville Superoperator Form}

Consider a unitary operator of the form \( U(x) = e^{-ixH} \), where \( H \) is a Hermitian operator. The corresponding quantum channel is:
\begin{equation}
    \check{U}(x, \rho) = U(x) \rho U^\dag(x). \label{eq:unitarychannel_appendix}
\end{equation}

Since \( H \) is a Hermitian operator, it admits a spectral decomposition \( H = V\Sigma V^\dag \), where \( \Sigma \) is a diagonal matrix of real eigenvalues \( \lambda_i \), and \( V \) is unitary. Therefore,
\begin{equation}
    U(x) = e^{-ixH} = e^{-ixV\Sigma V^\dagger} = V e^{-ix\Sigma} V^\dagger, \label{eq:spectralunitary_appendix}
\end{equation}
where the last equality follows from the property that for any function \( f \) and spectral decomposition \( A = V\Lambda V^\dagger \), we have \( f(A) = V f(\Lambda) V^\dagger \). Setting \( f(t) = e^{-ixt} \) and noting that \( e^{-ix\Sigma} = \text{diag}(e^{-ix\lambda_1}, e^{-ix\lambda_2}, \ldots, e^{-ix\lambda_n}) \), we obtain \( U(x) = V e^{-ix\Sigma} V^\dagger \).

Substituting into Eq.~\eqref{eq:unitarychannel_appendix}, we get:
\begin{equation}
    \check{U}(x, \rho) = V e^{-ix\Sigma} V^\dag \rho V e^{ix\Sigma} V^\dag. \label{eq:channeldecomp_appendix}
\end{equation}

Vectorising this using the identity \( \ket{ABC}\rangle = (C^T \otimes A)\ket{B}\rangle \), we obtain:
\begin{eqnarray}
    \hat{U}_{\text{Liouv}}(x) &=& (V e^{ix\Sigma} V^\dag)^T \otimes (V e^{-ix\Sigma} V^\dag) \nonumber \\
	&=& (V^\ast e^{ix\Sigma} V^T) \otimes (V e^{-ix\Sigma} V^\dag) \nonumber \\
	&=& (V^\ast \otimes V)(e^{ix\Sigma} \otimes e^{-ix\Sigma})(V^T \otimes V^\dag) \nonumber \\
    &=& V' e^{-ix\hat{\Sigma}} V'^\dag,
\end{eqnarray}
where \( V' = V^\ast \otimes V \) is unitary and the diagonal elements of \( \hat{\Sigma} \) are given by \( \lambda_j - \lambda_k \).

\subsubsection{Pauli Transfer Matrix Transformation}

To express the channel in the PTM form, we apply a basis transformation using the unitary \( T \) that maps computational basis operators to Pauli basis elements:
\begin{equation}
    T = \sum_{j,k} \ket{P_j}\rangle \langle\bra{c_k},
\end{equation}
where $P_j \in \mathcal{P}^{\otimes n}$ are Pauli basis elements and $c_k \in \{\ket{u}\bra{v} : u,v \in \{0,1\}^{\otimes n}\}$ are computational basis elements.

The PTM representation is then:
\begin{eqnarray}
    \hat{U}_{\text{PTM}}(x) &=& T \hat{U}_{\text{Liouv}}(x) T^\dag \nonumber \\
    &=& V'' e^{-ix\hat{\Sigma}} V''^\dag,
\end{eqnarray}
where \( V'' = T V' \) is also unitary.

This completes the derivation of the channel representations used in our analysis.

\subsection{Derivations for Ideal Quantum Models}
\label{appendix:ideal_model_derivations}

We provide here the full derivation of the Fourier coefficients and gradient expressions for the ideal quantum model introduced in \cref{sec:channel_representation}.

\subsubsection{Fourier Coefficients}

The coefficients $a_{\mathbf{j}, \mathbf{u}, \mathbf{v}}$ are given by:
\begin{widetext}
\begin{equation}
    a_{\mathbf{j}, \mathbf{u}, \mathbf{v}}(\boldsymbol{\theta}) = [\langle\bra{M}]_{u_{L+1}} \hat{W}^{(L+1)}_{u_{L+1} v_{L}} (\boldsymbol{\theta}) V_{v_{L} j_L} V^\dag_{j_L u_L} \hat{W}^{(L)}_{u_L v_{L-1}} (\boldsymbol{\theta}) \cdots V_{v_2 j_2} V^\dag_{j_2 u_2} \hat{W}^{(2)}_{u_2 v_1}(\boldsymbol{\theta}) V_{v_1 j_1} V^\dag_{j_1 u_1} \hat{W}^{(1)}_{u_1 v_0}(\boldsymbol{\theta}).
\end{equation}
\end{widetext}
where \( \mathbf{u} = (u_1, \ldots, u_{L+1}) \) and \( \mathbf{v} = (v_0, \ldots, v_{L}) \) index intermediary components of the PTM basis transformations.

\subsubsection{Gradient Expression}

Differentiating the model output with respect to a parameter \( \theta_i \), we obtain:
\begin{equation}
    \frac{\partial f}{\partial \theta_i}(x, \boldsymbol{\theta}) = \sum_{\omega} \frac{\partial c_{\omega}}{\partial \theta_i}(\boldsymbol{\theta}) e^{i\omega x},
\end{equation}
where,
\begin{equation}
    \frac{\partial c_{\omega}}{\partial \theta_i}(\boldsymbol{\theta}) = d \sum_{\substack{\mathbf{j} \\ \hat{\Lambda}_{\mathbf{j}} = \omega}} \sum_{\mathbf{u}, \mathbf{v}} \frac{\partial a_{\mathbf{j}, \mathbf{u}, \mathbf{v}}}{\partial \theta_i}(\boldsymbol{\theta}).
\end{equation}
which reveals how each frequency component contributes to the gradient landscape.

This completes the derivation of the ideal model’s Fourier structure and its parameter sensitivity.

\subsection{Derivations for Noisy Quantum Models}
\label{appendix:noisy_model_derivations}

This appendix provides detailed derivations for the expressions used in \cref{sec:results}, including gradient expressions under different noise models.

\subsubsection{Pauli Noise}

The gradient of the output with respect to a parameter \( \theta_i \) is:
\begin{equation}
    \frac{\partial \tilde{f}}{\partial \theta_i}(x, \boldsymbol{\theta}) = d \sum_{\mathbf{j}} e^{ix\Lambda_{\mathbf{j}}} \sum_{\mathbf{u}, \mathbf{v}} \frac{\partial \tilde{a}_{\mathbf{j}, \mathbf{u}, \mathbf{v}}}{\partial \theta_i}(\boldsymbol{\theta}),
\end{equation}
where
\begin{equation}
    \frac{\partial \tilde{a}_{\mathbf{j}, \mathbf{u}, \mathbf{v}}}{\partial \theta_i}(\boldsymbol{\theta}) = n_{\mathbf{u}} n_{\mathbf{v}} \frac{\partial a_{\mathbf{j}, \mathbf{u}, \mathbf{v}}}{\partial \theta_i}(\boldsymbol{\theta}). \label{eq:pauli_grad}
\end{equation}
Again, to directly connect this formulation to the Fourier coefficients, we express the noisy coefficients as:
\begin{equation}
    \frac{\partial \tilde{c}_\omega}{\partial \theta_i}(\boldsymbol{\theta}) = n_\omega (\boldsymbol{\theta})\frac{\partial c_\omega}{\partial \theta_i}(\boldsymbol{\theta}),
\end{equation}
This can also be seen by directly differentiating \cref{eq:noisecoeffrelation} using the chain rule and noting that $\frac{\partial n_\omega}{\partial \theta_i}(\boldsymbol{\theta}) = 0$ via the quotient rule. 

\subsubsection{Coherent Noise}

Coherent noise corresponds to unitary errors. Since unitary operators form a group under composition, the noisy circuit can be reparameterised:
\begin{equation}
    \tilde{f}(x, \boldsymbol{\theta}) = f(x, \boldsymbol{\theta}'),
\end{equation}
where \( \boldsymbol{\theta}' = h(\boldsymbol{\theta}) \) is a reparameterisation induced by the noise.

Using the chain rule, the gradient becomes:
\begin{equation}
    \frac{\partial \tilde{f}}{\partial \theta_i}(x, \boldsymbol{\theta}) = \frac{\partial f}{\partial \theta_i}(x, h(\boldsymbol{\theta})) \frac{\partial h}{\partial \theta_i}(\boldsymbol{\theta}).
\end{equation}

\subsubsection{Amplitude Damping}

Amplitude damping is a non-unital noise channel transforming $\ket{1}$ into the ground state $\ket{0}$ with probability $\gamma$. The coefficients $\tilde{a}_{\mathbf{j'}, \mathbf{u}, \mathbf{v}}$ include an additional term due to non-unital components. The full expression is:
\begin{widetext}
\begin{eqnarray}
    \tilde{a}_{\mathbf{j'}, \mathbf{u}, \mathbf{v}}(\boldsymbol{\theta}) = [\langle\bra{M}]_{u_{L+1}}&& \Big( 
	n_{u_{L+1}} \hat{W}^{(L+1)}_{u_{L+1} v_{L}} (\boldsymbol{\theta}) n_{v_L} V_{v_{L} j_L} V^\dag_{j_L u_L} n_{u_L} \cdots \nonumber \\
    && n_{v_{k+1}} V_{v_{k+1} j_{k+1}} V^\dag_{j_{k+1} u_{k+1}} n_{u_{k+1}} \hat{W}^{({k+1})}_{u_{k+1} v_k}(\boldsymbol{\theta}) n_{v_k} V_{v_k j_k} V^\dag_{j_k u_k} n_{u_k} \hat{W}^{(k)}_{u_k v_{k-1}}(\boldsymbol{\theta}) \nonumber \\
    && + n_{u_{L+1}} \hat{W}^{(L+1)}_{u_{L+1} v_{L}} (\boldsymbol{\theta}) n_{v_L} V_{v_{L} j_L} V^\dag_{j_L u_L} n_{u_L} \cdots \nonumber \\
    && n_{v_{k+1}} V_{v_{k+1} j_{k+1}} V^\dag_{j_{k+1} u_{k+1}} n_{u_{k+1}} \hat{W}^{({k+1})}_{u_{k+1} v_k}(\boldsymbol{\theta}) n_{v_k} V_{v_k j_k} V^\dag_{j_k u_k} \gamma_{u_k}\Big),
\end{eqnarray}
\end{widetext}

The gradients mirror the form of \cref{eq:noisyfourier}: 
\begin{equation}
    \frac{\partial \tilde{f}}{\partial \theta_i}(x, \boldsymbol{\theta}) = d \sum_{\mathbf{j'}} e^{ix\Lambda_{\mathbf{j'}}} \sum_{\mathbf{u}, \mathbf{v}} \frac{\partial \tilde{a}_{\mathbf{j'}, \mathbf{u}, \mathbf{v}}}{\partial \theta_i}(\boldsymbol{\theta}).
\end{equation}

\subsection{Resultant Channels from Twirling}
\label{appendix:channeltwirling}
We examine the effect of Pauli twirling on the PTM of a single qubit channel $\hat{R}$,
\begin{equation}\label{eq:paulitwirl}
    \hat{R} \longrightarrow \frac{1}{4} \sum_i \hat{P}_i^\dag \hat{R} \hat{P}_i,
\end{equation}
where $\hat{P}_i$ are the PTM representations of the Pauli operators,

\begin{eqnarray*}
    \hat{I} &=& \begin{bmatrix}
            1 & 0 & 0 & 0 \\
            0 & 1 & 0 & 0 \\
            0 & 0 & 1 & 0 \\
            0 & 0 & 0 & 1 
        \end{bmatrix}, \quad
    \hat{X} = \begin{bmatrix}
            1 & 0 & 0 & 0 \\
            0 & 1 & 0 & 0 \\
            0 & 0 & -1 & 0 \\
            0 & 0 & 0 & -1 
        \end{bmatrix}, \\
    \hat{Y} &=& \begin{bmatrix}
            1 & 0 & 0 & 0 \\
            0 & -1 & 0 & 0 \\
            0 & 0 & 1 & 0 \\
            0 & 0 & 0 & -1 
        \end{bmatrix}, \quad
    \hat{Z} = \begin{bmatrix}
            1 & 0 & 0 & 0 \\
            0 & -1 & 0 & 0 \\
            0 & 0 & -1 & 0 \\
            0 & 0 & 0 & 1 
        \end{bmatrix}.
\end{eqnarray*}
Applying an $\hat{X}$ twirl to channel $\hat{R}$ gives,
\begin{equation}
    \hat{X}\hat{R}\hat{X} = \begin{bmatrix}
         \hat{R}_{II} & \hat{R}_{IX} & -\hat{R}_{IY} & -\hat{R}_{IZ} \\
         \hat{R}_{XI} & \hat{R}_{XX} & -\hat{R}_{XY} & -\hat{R}_{XZ} \\
         -\hat{R}_{YI} & -\hat{R}_{YX} & \hat{R}_{YY} & \hat{R}_{YZ} \\
         -\hat{R}_{ZI} & -\hat{R}_{ZX} & \hat{R}_{ZY} & \hat{R}_{ZZ}
    \end{bmatrix}.
\end{equation}
Similarly, for $\hat{Y}$ and $\hat{Z}$ twirls,
\begin{eqnarray}
    \hat{Y}\hat{R}\hat{Y} &=& \begin{bmatrix}
         \hat{R}_{II} & -\hat{R}_{IX} & \hat{R}_{IY} & -\hat{R}_{IZ} \\
         -\hat{R}_{XI} & \hat{R}_{XX} & -\hat{R}_{XY} & \hat{R}_{XZ} \\
         \hat{R}_{YI} & -\hat{R}_{YX} & \hat{R}_{YY} & -\hat{R}_{YZ} \\
         -\hat{R}_{ZI} & \hat{R}_{ZX} & -\hat{R}_{ZY} & \hat{R}_{ZZ}
    \end{bmatrix},\\
    \hat{Z}\hat{R}\hat{Z} &=& \begin{bmatrix}
         \hat{R}_{II} & -\hat{R}_{IX} & -\hat{R}_{IY} & \hat{R}_{IZ} \\
         -\hat{R}_{XI} & \hat{R}_{XX} & \hat{R}_{XY} & -\hat{R}_{XZ} \\
         -\hat{R}_{YI} & \hat{R}_{YX} & \hat{R}_{YY} & -\hat{R}_{YZ} \\
         \hat{R}_{ZI} & -\hat{R}_{ZX} & -\hat{R}_{ZY} & \hat{R}_{ZZ}
    \end{bmatrix}.
\end{eqnarray}
Computing the resultant channel as per \cref{eq:paulitwirl}, 
\begin{equation}
    \frac{1}{4} \sum_i \hat{P}_i^\dag \hat{R} \hat{P}_i = \begin{bmatrix}
         \hat{R}_{II} & 0 & 0 & 0 \\
         0 & \hat{R}_{XX} & 0 & 0 \\
         0 & 0 & \hat{R}_{YY} & 0 \\
         0 & 0 & 0 & \hat{R}_{ZZ}
    \end{bmatrix}.
\end{equation}
Thus, Pauli twirling zeros all off-diagonal elements of the PTM while preserving its diagonal entries.

For Clifford twirling, the resultant PTM is,
\begin{equation}
    \frac{1}{K} \sum_{k=1}^K \hat{C_k}^\dag \hat{R} \hat{C_k} = \begin{bmatrix}
         \hat{R}_{II} & 0 & 0 & 0 \\
         0 & \overline{\lambda} & 0 & 0 \\
         0 & 0 & \overline{\lambda} & 0 \\
         0 & 0 & 0 & \overline{\lambda}
    \end{bmatrix},
\end{equation}
where $\hat{C}_k$ are PTM representations of Clifford group elements, and $\overline{\lambda} = (\hat{R}_{XX} + \hat{R}_{YY} + \hat{R}_{ZZ})/3$.

\subsubsection{Amplitude Damping}
The PTM for a single qubit amplitude damping channel is,
\begin{equation}
    \hat{R}_\text{AD} = \begin{bmatrix}
        1 & 0 & 0 & 0 \\
        0 & \sqrt{1-\gamma} & 0 & 0 \\
        0 & 0 & \sqrt{1-\gamma} & 0 \\
        \gamma & 0 & 0 & 1-\gamma
    \end{bmatrix}.
\end{equation}

When Pauli-twirled,
\begin{equation}
    \frac{1}{4} \sum_i \hat{P}_i^\dag \hat{R}_\text{AD} \hat{P}_i = \begin{bmatrix}
        1 & 0 & 0 & 0 \\
        0 & \sqrt{1-\gamma} & 0 & 0 \\
        0 & 0 & \sqrt{1-\gamma} & 0 \\
        0 & 0 & 0 & 1-\gamma
    \end{bmatrix}.
\end{equation}

A Pauli noise channel that has the distribution $\{ ('I', 1-p_X-p_Y-p_Z),\; ('X', p_X),\; ('Y', p_Y),\; ('Z', p_Z) \}$, has a PTM given by,
\begin{equation}
    \hat{R}_\text{Pauli} = \left[\begin{smallmatrix}
        1 & 0 & 0 & 0 \\
        0 & 1-2p_Y-2p_Z & 0 & 0 \\
        0 & 0 & 1-2p_X-2p_Z & 0 \\
        0 & 0 & 0 & 1-2p_X-p_Y
    \end{smallmatrix}\right].
\end{equation}
Therefore, to find a Pauli noise channel of this form that is equivalent to Pauli-twirled amplitude damping, we may solve simultaneous equations given by the diagonal elements of the matrix equality,
\begin{equation}
    \frac{1}{4} \sum_i \hat{P}_i^\dag \hat{R}_\text{AD} \hat{P}_i = \hat{R}_\text{Pauli}.
\end{equation}

\section{Supplementary Numerical Investigations}
\label{appendix:B}

\subsection{Simulation Details}
\label{appendix:sim_details}
\subsubsection{Data Reuploading Model Configuration}
In all data reuploading circuit simulations in the main text, circuits are initialised on the ground state, $\ket{0}$, and measured in the $Z$ basis. Each trainable block \( W^l(\boldsymbol{\theta}) \) is implemented as a sequence of three single-qubit rotations: a Z-rotation by \(\theta_i\), followed by a Y-rotation by \(\theta_{i+1}\), and another Z-rotation by \(\theta_{i+2}\). Each parameter \(\theta_i\) is used exactly once across all instances of \( W^l \). The data-encoding gate \( S(x) \) is realised as a single X-rotation with angle \( x \) (in radians). In \cref{fig:exprange}(a) the number of repeated layers $L = 2$ and \cref{fig:grad_decay}. The examples in \cref{fig:exprange}(b) use $L=3$. Training data used for \cref{fig:exprange} consist of 250 input points \( x \in [-2\pi, 2\pi] \), with corresponding target values sampled from a truncated Fourier series with randomised coefficients. Each target function has coefficients drawn uniformly from a bounded range. The model is trained using the ADAM optimiser to minimise mean squared error (MSE), with data shuffled and processed in batches of 25. Refer to \cref{tab:hyperparams} for the hyperparameters used by the ADAM optimiser.

\begin{table}[b]
\caption{\label{tab:hyperparams}%
ADAM optimiser hyperparameters used in data reuploading circuit training for each circuit depth.
}
\begin{ruledtabular}
\begin{tabular}{c|cccc}
\textrm{Depth ($L$)}&
\textrm{Learning Rate}&
\textrm{$\beta_1$}&
\textrm{$\beta_2$}&
\textrm{Max Steps}\\
\colrule
1 & 0.30 & 0.45 & 0.90 & 100\\
2 & 0.20 & 0.45 & 0.95 & 150\\
3 & 0.08 & 0.70 & 0.925 & 350\\
4 & 0.10 & 0.725 & 0.99 & 500\\
5 & 0.06 & 0.725 & 0.925 & 1250\\
6 & 0.03 & 0.75 & 0.95 & 3000\\
\end{tabular}
\end{ruledtabular}
\end{table}




\subsection{Gradient Distributions under Twirled Amplitude Damping}
\label{appendix:pauli_loss}
\begin{figure*}
\includegraphics{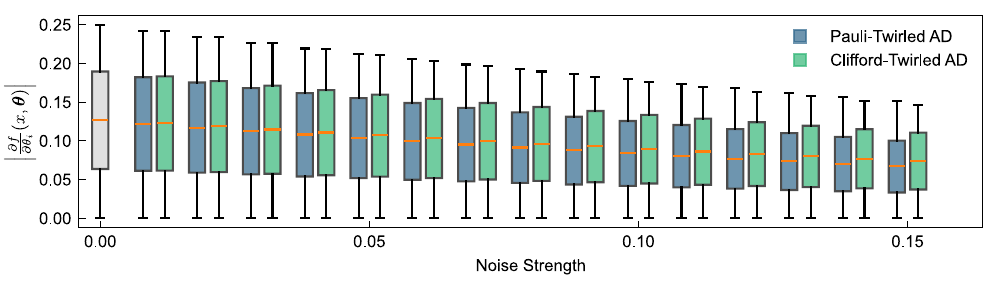}
\caption{\label{fig:grad_decay_twirl}
Effect of noise on gradient magnitudes during training. 
Box plots show the distribution of absolute gradient values, \(\left| \frac{\partial \tilde{f}}{\partial \theta_i} \right|\), for a two-layer quantum circuit subjected to varying strengths of Pauli-twirled and Clifford-twirled amplitude damping noise. Each distribution is computed from 10{,}000 randomly sampled parameter sets and input values. At higher noise strengths, Pauli-twirled noise yields slightly larger maximum gradients; however, the inner quartiles are consistently higher under Clifford-twirled noise, indicating a more broadly robust gradient landscape for optimisation.}
\end{figure*}

To complement the analysis in the main text, we examine the effect of twirled amplitude damping noise on gradient magnitudes during training. Specifically, we compare Pauli-twirled amplitude damping with Clifford-twirled amplitude damping, using the same experimental setup described in \cref{fig:grad_decay_twirl}.

For each of 10{,}000 randomly sampled parameter sets \(\boldsymbol{\theta} = \{\theta_1, \ldots, \theta_9\}\), we randomly select one parameter \(\theta_i\) and one input \(x \sim \mathcal{U}(0, 2\pi)\). The absolute gradient \(\left| \frac{\partial \tilde{f}}{\partial \theta_i}(x, \boldsymbol{\theta}) \right|\) is computed analytically for each noise setting. These distributions are visualised as box plots to characterise the trainability of the circuit under different noise models.

As shown in \cref{fig:grad_decay_twirl}, Pauli-twirled noise had a greater suppression of gradient magnitudes, indicating a flatter loss landscape. These results are consistent with the findings in \cref{appendix:asym_pauli}, where it was shown that, for this circuit configuration, the optimiser is most effective at mitigating the impact of $Z$-axis Pauli noise, while $X$-axis noise proves most detrimental. Since Pauli-twirled amplitude damping introduces stronger distortions along the $X$ and $Y$ axes and weaker distortions along the $Z$ axis, the observed suppression of gradients under Pauli twirling aligns with the circuit’s heightened sensitivity to $X$-axis noise.

\subsection{Biased Pauli Noise}
\label{appendix:asym_pauli}

\begin{figure}[]
\includegraphics[]{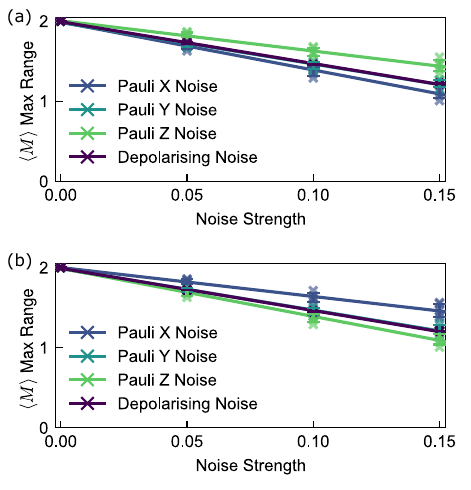}
\caption{\label{fig:exprange_pauli}
Effect of biased Pauli noise on model expressivity.
This figure shows the output range of a single-qubit data reuploading circuit under various Pauli noise channels. Expressivity is evaluated by gradually increasing the amplitude of target functions until the model fails to fit the data. The comparison highlights differences between biased Pauli noise, where noise acts along a single axis, and fully symmetric depolarising noise, which corresponds to the Clifford-twirled versions of each Pauli noise channel at equal noise strength.
(a) Using the circuit architecture described in \cref{appendix:sim_details}, we find that the optimiser most effectively mitigates the impact of $Z$-axis Pauli noise. In contrast, $X$-axis noise has the most detrimental effect on model performance, while $Y$-axis noise behaves similarly to depolarising noise.
(b) In an alternative circuit configuration, we swap $Z$ rotation gates with $X$ rotation gates and vice versa, initialise in the $X$ basis, and measure in the $X$ basis. This reversal of rotation axes leads to a corresponding reversal in the observed effects of $Z$ and $X$ Pauli noise, highlighting the sensitivity of noise resilience to the specific structure of the quantum circuit.}
\end{figure}

In this section, we extend our investigation beyond the twirled amplitude noise considered in the main text, focusing instead on the effects of biased Pauli noise channels on supervised learning performance. Here, biased refers to the non-uniformity of the noise; specifically, noise that acts along a single Pauli axis ($X$, $Y$, or $Z$), as opposed to depolarising noise, which distributes uniformly across all three axes.

To assess the impact of biased noise on model expressivity, as found in the main text, we measure the output range of a single-qubit data reuploading circuit under different noise conditions. This is done by gradually increasing the amplitude of target functions until the model fails to fit the data, thereby identifying the threshold at which expressivity breaks down. Figure~\ref{fig:exprange_pauli} presents these results for two circuit configurations.

In panel (a), we use the standard circuit architecture described in \cref{appendix:sim_details}. Under this setup, the optimiser is most effective at mitigating the impact of $Z$-axis Pauli noise, while $X$-axis noise proves most detrimental. Finally, $Y$-axis noise yields behaviour nearly indistinguishable from depolarising noise. These observations help explain the performance gap seen in \cref{fig:exprange} of the main text. Pauli-twirled amplitude damping, composed of stronger $X$ and $Y$ components and weaker $Z$, results in poorer expressivity compared to Clifford-twirled amplitude damping. The latter more evenly distributes noise and therefore avoids amplifying the most disruptive components for this circuit.

Panel (b) explores an alternative circuit configuration in which $Z$ rotation gates are swapped with $X$ rotation gates and vice versa. Additionally, the circuit is initialised and measured in the $X$ basis. This structural modification results in a reversal of the observed effects of $Z$ and $X$ Pauli noise, highlighting the sensitivity of noise resilience to the specific structure of the quantum circuit.

These findings suggest that circuit design plays a critical role in shaping the model's robustness to biased noise, and that certain configurations may inherently amplify or suppress the impact of non-uniform noise channels. This insight could inform future strategies for noise-aware circuit optimisation in quantum machine learning.

\subsection{VQE: Reversing Directional Noise}
\label{appendix:VQE_ad_reverse}
\begin{figure}[]
\includegraphics[]{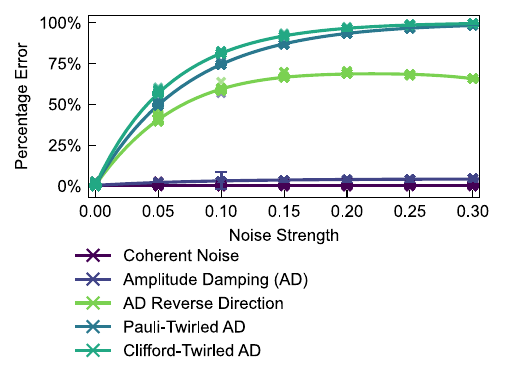}
\caption{\label{fig:VQE_error_adr}
Impact of reversed amplitude damping on VQE performance. 
This figure extends the results shown in \cref{fig:VQE_error} by including a non-physical noise model. In this case, amplitude damping occurs in the reverse direction, exciting \(\ket{0}\) to \(\ket{1}\) with probability \(\gamma\), which corresponds to the noise strength shown on the x-axis. While not physically realistic, this reversed noise channel leads to significantly worse performance than standard amplitude damping, highlighting the strong dependence of model behaviour on the structure and directionality of noise.}
\end{figure}

To further explore the sensitivity of variational quantum algorithms to the structure of noise, we consider a non-physical variant of amplitude damping in which the direction of decay is reversed. Specifically, rather than relaxing from the excited state \(\ket{1}\) to the ground state \(\ket{0}\), this reversed channel excites \(\ket{0}\) to \(\ket{1}\) with probability \(\gamma\), where \(\gamma\) corresponds to the noise strength shown on the x-axis of \cref{fig:VQE_error_adr}.

While this reversed amplitude damping channel does not correspond to any physical process, it serves as a useful probe of how directional biases in noise interact with circuit structure. As shown in \cref{fig:VQE_error_adr}, this reversed noise model leads to significantly worse performance than standard amplitude damping, with the optimiser failing to recover accurate ground state energies even at moderate noise strengths. 

Despite this degradation in performance, the reversed channel still outperforms its Pauli- and Clifford-twirled counterparts. Due to the symmetrisation, both twirls of the reversed channel are mathematically equivalent to the twirls of standard amplitude damping. Therefore, twirling the reversed channel further erodes performance, despite its already non-physical and poorly aligned structure.

These results reinforce the broader conclusion that symmetrising noise can be more harmful than preserving its natural biases. While biases alone do not guarantee good performance, maintaining directional structure can help retain features that the optimiser can exploit. In contrast, twirling obscures this structure and introduces uniformly disruptive components, ultimately degrading both expressivity and trainability.

\end{document}